\documentclass[%
 reprint,
 amsmath,amssymb,
 aps,
]{revtex4-2}

\usepackage{graphicx}
\usepackage{dcolumn}
\usepackage{bm}
\usepackage{amsfonts} 
\usepackage[T1]{fontenc} 
\usepackage{orcidlink}
\usepackage{mathrsfs}
\usepackage{amsmath}

\begin{document}

\preprint{APS/123-QED}

\title[Spin torque driven mode hybridization and band engineering in nanopatterned magnonic crystals]{Spin torque driven mode hybridization and band engineering in nanopatterned magnonic crystals}
\author{Nikhil Kumar, \orcidlink{0000-0003-3213-505X}}
\email{nikhilkumarcs@nitc.ac.in}

\affiliation{Electronics and Communication Engineering Department, National Institute of Technology Calicut, 673 601 India}

\date{\today}

\begin{abstract}
Spin wave propagation and its control at the nanoscale are at the heart of the development of reconfigurable magnonic and spintronic devices. Here, a tunable mode coupling and band hybridisation is demonstrated in a nanopatterned magnonic crystal through the effective utilisation
of inhomogeneous current-induced spin torque. The system consists of a Permalloy thin film interfaced with a heavy metal and patterned
with a two dimensional array of circular Co nanodots, thereby forming a bicomponent magnonic crystal with spatially induced magnetic and
spin-torque parameters. The spin wave band structures and modes are investigated using the plane wave method(PWM). By solving the Linearised
Landau-Lifshitz equation with a field-like torque term in the reciprocal lattice, a full magnonic band gap is obtained. Injection of current
through the heavy-metal layer induces inhomogeneous spin-orbit coupling in magnonic crystals, resulting in a periodic perturbation of the magnonic frequency and enabling dynamic control of spin-wave dispersion. A prominent avoided crossing between localised modes and propagating Damon--Eshbach--type modes at a finite wave vector is observed. The presence of a periodic effective spin-torque term leads to band
deformation, enhanced nanoscale mode mixing, and a tunable hybridisation gap. Eigenmodes reveal the transition from confined standing modes
to propagating modes, and nanoscale mode exchange and band repulsion, with spin torque term control, determine the position and magnitude
of the anticrossing in the spin wave dispersion. Eigenmodes reveal the transition from confined standing modes to propagating modes, and mode exchange and band repulsion, with spin torque term control, determine the position and magnitude of the anticrossing in the band structure. These results show that spin torque effectively controls
the hybrid magnonic states, enabling tunable mode conversion and enhanced control over spin-wave functionality in magnonic crystals and spintronic
devices. The anticrossing position and the hybridisation gap are effectively controlled by applied current, enabling dynamic reconfiguration of the magnonic state. 
\end{abstract}

\maketitle

The collective excitations of magnetization in ordered magnetic materials, the so-called spin waves, have attracted significant attention as
potential carriers for next-generation low-power computing and nanoscale microwave technologies because of their shorter wavelengths in the
nanoscale regime, coherent wave nature, and much lower Joule heating compared to charge-based electronics\cite{Intro_1_kruglyak2010magnonics,intro_2_lenk2011building,Intro_3_chumak2015magnon}.
Magnonics uses spin-wave propagation and interference to transfer information. It is now seen as a promising area of research for wave-based
computing, neuromorphic systems, and coherent microwave devices\cite{Intro_4_demokritov2012magnonics,Intro_5_khitun2011non,Intro_6_chumak2017magnonic}.
Periodic magnetic systems, known as magnonic crystals, enable artificial control of spin-wave dispersion through internal-field modulation,
Bragg scattering, and geometrical confinement\cite{Intro_7_nikitov2001spin,Intro_8_bauer2012spin,Intro_9_krawczyk2014review,Intro_10_grundler2015reconfigurable}. These periodic magnonic lattices have adjustable magnonic band structures,
allow for directional spin-wave movement, support both localized and traveling modes, and create frequency-selective band gaps\cite{Intro_11_mruczkiewicz2017spin,Intro_12_gruszecki2015influence,Intro_13_krawczyk2008plane,Intro_14_klos2013magnonic}.
Two-dimensional bicomponent magnonic crystals (2D BMCs), composed of magnetic dots within a continuous ferromagnetic matrix, are of
interest because they support strong dipolar-exchange interactions,
mode localization, and complex Bloch-wave dynamics that periodically alter magnetic properties\cite{Intro_15_rychly2015magnonic,Intro_16_wagner2016magnetic,Intro_17_sadovnikov2015magnonic,Intro_18_tacchi2012forbidden}.
Researchers have found that structural periodicity, magnetic anisotropy,
interfacial interactions, and material contrast can be used to control spin-wave band engineering\cite{Intro_19_neusser2009magnonics,Intro_20_mruczkiewicz2016collective,Intro_21_ciubotaru2016all,Intro_22_khitun2010magnonic}.
In addition, magnonic filters, waveguides, logic elements, and reconfigurable nanoscale microwave components can be created by controlling the spin
wave band structure at the nanoscale\cite{Intro_23_chumak2014magnon,intro_24_schneider2008realization,intro_25_fulara2019spin,Intro_26_khitun2005nano}.
Even with recent progress, most magnonic crystals work as passive systems and their dynamic behavior stays the same after they are made. This makes them less useful for adaptive or programmable magnonic technologies\cite{Intro_27_klingler2014design,Intro_28_chumak2022advances,Intro_29_grundler2016nanomagnonics}.

Recent advances in spintronics and spin-orbit torque physics have established a novel platform for the active electric control of spin
wave propagation and magnetization dynamics in nanoscale magnetic heterostructures\cite{Intro_30_miron2011perpendicular,Intro_31_liu2012spin,Intro_32_hoffmann2013spin,Intro_33_tserkovnyak2005spin}.
Injection of current into heavy-metal/ferromagnet bilayers induces spin accumulation through strong spin-orbit torque coupling and the
spin Hall effect. This process generates an effective torque, enabling manipulation of magnetic dynamics on ultrafast timescales\cite{Intro_34_sinova2015spin,Intro_35_manchon2019current,Intro_36_demidov2012nanooscillator,Intro_37_hamadeh2014control}. Spin transfer torque and spin-orbit torque mechanisms have opened
up broad research on current-driven magnetization switching, spin-wave amplification, and nanoscale microwave nanoscillators\cite{Intro_38_Demidov2014SHNO,Intro_39_Brataas2012Torques,Intro_40_Demidov2016Coherent,Intro_41_Divinskiy2018Magnonics,Intro_42_Cros2013STNO}.
In recent developments, electrically driven spin-wave excitation and tunable magnon propagation have been identified as promising strategies for low-power coherent signal processing and active functionality\cite{Intro_43_Urazhdin2014Nanomagnonic,Intro_44_Demidov2015Waveguides,Intro_45_Anane2020STNO,Intro_46_Demidov2020SOTMagnonics}.
Incorporating the spin torque mechanism into periodic magnetic systems enables dynamic modification of spin-wave band structures, mode coupling, and reciprocal-space interactions\cite{Intro_47_Yu2014Damping,Intro_48_Li2022Hybrid,Intro_49_Li2019Coherent}.
Active manipulation of magnonic eigenstates is achieved through current-induced spin torque, which significantly affects damping, phase coherence, and nonlinear mode interactions\cite{Intro_50_Lachance2019Hybrid,Intro_51_Zhang2014Strongly,intro_52_Tabuchi2014Hybridizing,Intro_53_Grundler2022Quantum}.
Magnons inherently function within the gigahertz frequency range, which aligns with the operational frequencies of superconducting quantum
circuits and cavity magnonic architectures. Consequently, electrically controllable magnonic systems represent a promising platform for hybrid
quantum technologies and coherent microwave applications\cite{Intro_54_Osada2016Optomagnonics,Intro_55_Walker1957Magnetostatic,Intro_56_Damon1961Magnetostatic,Intro_57_Kittel2005SolidState}.
Consequently, the dynamic tuning of magnon hybridization and intermodal coupling through electrical signals represents a central objective
in advancing quantum magnonic and coherent wave-based computing systems\cite{Intro_58_Holstein1940Field,Intro_59_Bloch1930Ferromagnetismus,Intro_60_Slonczewski1996Current}. Nevertheless, the influence of periodically modulated spin torque on avoided crossing behavior, band hybridization, and spin-wave mode
evolution in bicomponent two-dimensional magnonic crystals has not been thoroughly investigated\cite{Intro_61_Berger1996Emission,Intro_62_Demokritov2006BEC,Intro_63_Serga2010YIG}.

This study extensively investigates electrically tunable spin wave band engineering in nanopatterned bicomponent magnonic crystals driven
by inhomogeneous current-induced spin-orbit torque. The proposed device comprises a Permalloy thin film interfaced with a heavy-metal layer
and patterned with a two-dimensional periodic array of circular cobalt nanodots, resulting in simultaneous modulation of both magnetic and
spin-torque parameters. A systematic analysis of the influence of periodic effective spin torque on the magnonic band structure and spin-wave eigenmodes is conducted using the plane-wave method with the linearised Landau--Lifshitz equation incorporating a field-like spin-torque term in reciprocal space. Current-induced spin-torque perturbation dynamically modifies spin-wave dispersion and generates tunable avoided crossings between localised magnon modes and propagating Damon--Eshbach-like modes at finite wave vectors. Periodic injection of effective spin-orbit torque results in electrically controllable hybridisation gaps, increased mode coupling, and significant reconstruction
of magnonic eigenstates. Eigenmode analysis shows that increasing spin-torque strength shifts the system from confined standing-wave states to propagating Bloch-like modes, highlighting the dynamic evolution of spin-wave propagation. The position and magnitude of the anticrossing
gap can be adjusted through current injection, confirming spin torque as an effective tool for programmable magnonic band engineering. These
findings indicate strong potential for reconfigurable magnonic devices, adaptive microwave technologies, coherent wave-based information processing, and future quantum magnonic platforms\cite{Intro_64_Wang2020Neuromorphic,Intro_65_Wang2021Coherent,Intro_66_Chumak2022Unconventional,Intro_67_Grundler2020Reconfigurable}.

\section*{Method of Calculations}

The plane wave method (PWM) is used to calculate spin-wave dispersions and their magnetisation distributions. It works for systems with discrete
translational symmetry, such as photonic, phononic, and magnonic crystals (MCs). The method also applies to electronic superlattices\cite{joannopoulos2008,ashcroft1976,prather2009,krawczyk2008}. It is suitable for any lattice type, different periodicities, dimensions,
and any shape of scattering centres\cite{tiwari2010,klos2012,mamica2012}.
It gives a complete spectrum of spin-wave excitations in the structures we model. The method can also calculate defect states in superlattices\cite{yang2012}. These approaches help us develop a perturbation method for MCS, similar to those used in solid-state theory \cite{ashcroft1976}, photonics \cite{yariv2003}, and phononics \cite{kichin2012}.

Spin-orbit coupling is extensively used to convert charge current to spin current in various spintronic devices\cite{dyakonov1971current,jungwirth2012spin,tartakovskaya2000,murakami2003dissipationless,valenzuela2006direct}. When the spin current from spin-orbit coupling is absorbed by ferromagnets,
it exerts a spin-orbit torque (SOT) on them. From the viewpoint of nanoscale physics and its applications, it is crucially important
to understand the detailed SOT characteristics and their magnetisation dynamics in ferromagnetic layers. The SOT is commonly decomposed into
two mutually orthogonal vector components, the damping-like torque(DLT) and field-like torque (FLT). SOT-induced magnetisation dynamics, including the DLT and FLT terms, is described by the Landau-Lifshitz-Gilbert equation. The SOT-induced magnetization dynamics, including the DLT and FLT terms, is described by the Landau-Lifshitz-Gilbert equation

\begin{equation}
	\begin{split}
		\frac{\partial \hat{\mathbf{m}}}{\partial t}
		=&
		-\gamma\mu_{0}
		\left[
		\hat{\mathbf{m}}
		\times
		\mathbf{H}_{\mathrm{eff}}
		\right]
		+
		\alpha
		\left[
		\hat{\mathbf{m}}
		\times
		\frac{\partial \hat{\mathbf{m}}}{\partial t}
		\right]
		\\
		&
		+
		\gamma\tau_{\mathrm{d}}
		\hat{\mathbf{m}}
		\times
		\left[
		\hat{\mathbf{m}}
		\times
		\hat{\mathbf{y}}
		\right]
		+
		\gamma\tau_{\mathrm{f}}
		\hat{\mathbf{m}}
		\times
		\hat{\mathbf{z}}
		\label{eq:1}
	\end{split}
\end{equation}

where $\gamma$ is the gyromagnetic ratio, $\hat{\mathbf{m}}$ is the unit vector along the magnetization, $\mu_{0}\mathbf{H}_{\text{eff}}$
is the effective fields, including the external, anisotropic, magnetostatic, and exchange fields, $\alpha$ is the damping constant, $\tau_{\text{d}}=(\hbar/2e)(J/M_{s}d)c^{||}$
and $\tau_{\text{f}}=(\hbar/2e)(J/M_{s}d)c^{\perp}$ represents the magnitude of DLT and FLT, respectively, where $J$ is the current
density, $M_{\text{s}}$ is the saturation magnetization, $d$ is the thickness of ferromagnet, $c^{||}$ and $c^{\perp}$ are the DLT
and FLT efficiencies, respectively, and $\mathbf{\hat{z}}$ is the unit vector perpendicular to both current direction and the inversion asymmetry direction(that is thickness direction; see the coordinate system in Fig. 1. From Eq. 1, the two torque components affect the
magnetization dynamics in different ways: The DLT directs the magnetization toward the $z$ axis, whereas the FLT induces magnetization precessions
around the $z$ axis. To focus on the role of the field-like term in controlling spin wave dispersion and mode hybridisation, the effects
of Gilbert damping and the damping-like contribution are neglected in Eq. \ref{eq:1}. These terms mainly affect dissipation and linewidth, not the real part of the eigenfrequency. Leaving them out does not
change the spin wave dispersions. The Eq.(\ref{eq:1}) reduces to
\begin{equation}
	\cfrac{\partial\hat{\mathbf{m}}}{\partial t}=-\gamma\mu_{0}\left[\hat{\mathbf{m}}\times\mathbf{H}_{\text{eff}}\right]+\gamma\tau_{\text{f}}\hat{\mathbf{m}}\times\hat{\mathbf{z}}\label{eq:2}
\end{equation}

For PWM calculations, we assume that each magnetic material in the bicomponent magnonic crystal (BMC), shown in Fig. 1(b), has a uniform
static magnetisation aligned with the external magnetic field. This assumption allows us to use the linear approximation and a global coordinate system, in which the $y$- and $z$-axes define the plane of periodicity, and the $x$-axis is perpendicular to the surface of the magnonic crystal. In the case of SWs in the linear regime,
the component of the magnetization vector parallel to the static magnetic field is constant in time t (in this study, $\mathbf{H}_{0}$ is always
assumed to be oriented along the $z$-axis), and its magnitude is much greater than that of the perpendicular components: $|\mathbf{m}\left(\mathbf{r},t\right)|<<M_{z}\left(\mathbf{r}\right)$,
with $M\left(\mathbf{r},t\right)=M_{z}\left(\mathbf{r}\right)\hat{z}+\mathbf{m}\left(\boldsymbol{r},t\right)\left[r=\left(x,y,z\right)\right]$.
Thus, in linear approximation we neglect all terms with squared $\boldsymbol{m\left(r,t\right)}$
and $\boldsymbol{h_{\text{ms}}}\left(r,t\right)$. In linear approximations, all term with squared and its higher order $\text{\textbf{m}}\left(r,t\right)$
and $\boldsymbol{h_{\text{ms}}}\left(r,t\right)$ is neglected and assume $M_{z}=M_{s}$, $M_{s}$ being saturation magnetizations. For
monochromatic spin waves, $\mathbf{m}\left(\mathbf{r},t\right)\sim\text{exp\ensuremath{\left(i\omega t\right)}}\left(\text{\ensuremath{\omega}}\text{being the wave frequency}\right)$.

\begin{figure}[!ht]
 \includegraphics[width=1\columnwidth]{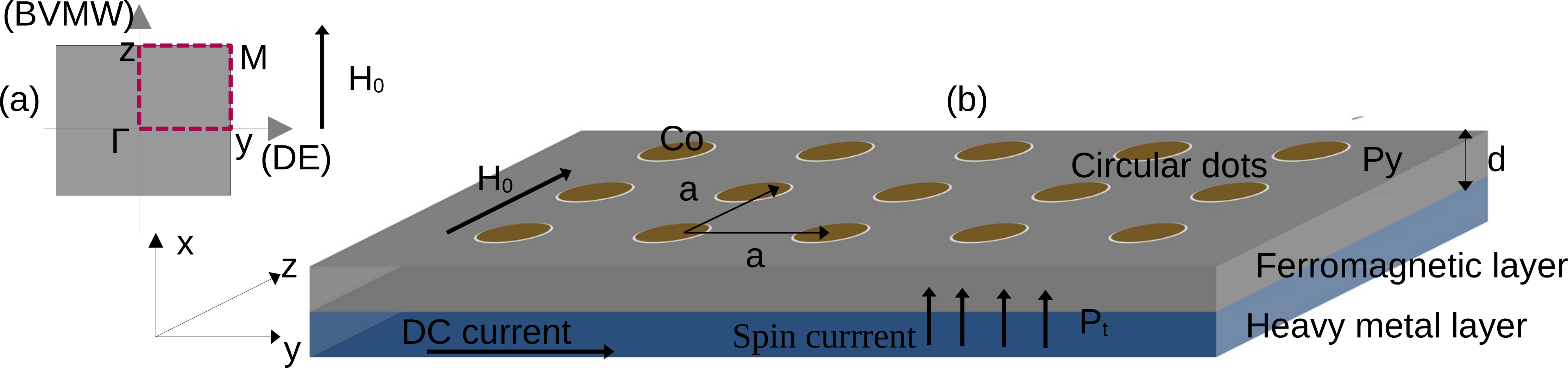}
	\caption{(a)First BZ for the structure shown in (b), with indicate high-symmetry
		points, $\Gamma$, $Y$, $M$ and $Z$ in the centre and at the border
		of the first BZ. There are two directions of the wave vector $k$
		along the bias magnetic field and perpendicular to it, i.e., the backward
		volume magnetostatic wave(BVMW) and Damon--Eshbach (DE) geometry,
		respectively. (b)Permalloy thin film interfaced with a heavy metal
		and patterned with a two-dimensional array of circular Co nanodots,
		thereby forming a bicomponent magnonic crystal(BMC) with spatially
		induced magnetic and spin-torque parameters. The thickness of the
		bicomponent magnonic crystal is $d$, the lattice constant is $a$.
		The magnetic field $H_{0}$ is always oriented along the $z$-axis.
		The $x$-axis is normal to the 2D MC plane. }\label{fig:(a)First-BZ-for}
\end{figure}

The effective magnetic field $\boldsymbol{\text{H}_{\text{eff}}}$
composed of the uniform magnetic field, the exchange field and the magnetostatic field $\boldsymbol{\text{H}_{\text{ms}}}$. The exchange
field and the magnetostatic field both vary with space and time. The exchange field is derived directly from the exchange-energy functional
under the linear approximation\cite{krawczyk2008,krawczyk2012formulation}. In magnetically nonhomogeneous material both the spatial nonhomogeneity of the exchange stiffness constant $A\left(\mathbf{r}\right)$, and
that of the spontaneous magnetization $M_{s}\left(\mathbf{r}\right)$ must be taken into account, leads to the following equation 
\begin{equation}
	\mathbf{H}_{\text{ex}}\left(\mathbf{r},t\right)=\left[\mathbf{\nabla\cdot\lambda^{2}_{\text{ex}}}\left(\mathbf{r,}t\right)\nabla\right]\mathbf{M}\left(\mathbf{r},t\right),\text{where \ensuremath{\lambda_{\text{ex}}=\sqrt{\frac{2A}{\mu_{0}M^{2}_{s}}}}}\label{eq:3}
\end{equation}
\noindent$\lambda_{\text{ex}}$ is the exchange length, $A$ is the exchange stiffness constant, and $M_{s}$ denotes the spontaneous (saturation magnetization). The term $\tau_{\text{f}}$ in Eq. (\ref{eq:1}) denotes the magnitude of field like torque(FLT) term and the corresponding spin torque term acts an effective fields. The magnetostatic fields is decomposed into a static and dynamic component, $\boldsymbol{H}_{\text{ms}}\left(\boldsymbol{r}\right)$ and $\boldsymbol{h}_{\text{ms}}\left(\boldsymbol{r},t\right)$, respectively. The time dependence of the dynamic magnetostatic field has the same
form as that of the dynamic component of the magnetisation vector:
$\boldsymbol{h}_{\text{ms}}\left(\boldsymbol{r},t\right)=\boldsymbol{h}_{\text{ms}}\left(\boldsymbol{r}\right)e^{i2\pi ft}$.
In the MCs assumed here (Fig. \ref{fig:(a)First-BZ-for}-b) the material
parameters: $M_{s}$ and $l_{\text{ex}}$, are periodic functions of the in-plane position vector $\boldsymbol{r}_{||}=\left(y,z\right)$,
with a period equal to the lattice vector \textbf{a} with lattice constant a. The Bloch's theorem asserts that a solution of a differential
equation with periodic coefficients can be represented as a product of a plane-wave envelope function and a periodic Bloch function: 
\begin{equation}
	\mathbf{m}\left(\mathbf{r}\right)=\mathbf{m}_{\mathbf{k}}\left(\mathbf{r}\right)\text{e}^{i\mathbf{k}\cdot\mathbf{r}}=\underset{\mathbf{G}}{\sum}\mathbf{m}_{\mathbf{k}}\left(\mathbf{G}\right)\text{e}^{i\left(\mathbf{k}+\mathbf{G}\right)\cdot\mathbf{r}}\label{eq:4}
\end{equation}
\noindent where $\mathbf{m}_{\mathbf{k}}\left(\mathbf{r}+a\right)=\mathbf{m}_{\mathbf{k}}\left(\mathbf{r}\right)$,
where $\mathbf{k}$ is the wave vector in the first Brillouin zone
and $\mathbf{G}$ denotes a reciprocal lattice vector. The Fourier
transformation to map the periodic function $M_{s}$ $l^{2}_{\text{ex}}$and
$\tau_{\text{f}}$ to the reciprocal space gives 
\begin{equation}
	\begin{array}{c}
		M_{s}\left(\mathbf{r}\right)=\underset{\mathbf{G}}{\sum}M_{s}\left(\mathbf{G}\right)\text{e}^{i\mathbf{G}\cdot\mathbf{r}},\text{\ensuremath{\lambda^{2}_{\text{ex}}\left(\mathbf{r}\right)=\underset{\mathbf{G}}{\sum}\lambda^{2}_{\text{ex}}\left(\mathbf{G}\right)\text{e}^{i\mathbf{G}\cdot\mathbf{r}}},}\\
		\\\tau_{\text{f}}\left(\boldsymbol{r}\right)=\underset{\mathbf{G}}{\sum}\tau_{\text{f}}\left(\mathbf{G}\right)\text{e}^{i\mathbf{G}\cdot\mathbf{r}}
	\end{array}\label{eq:5}
\end{equation}

\noindent In the case of circular dots the Fourier components of saturation magnetization $M_{s}\left(\mathbf{r}\right),$ the squared exchange length $\lambda^{2}_{\text{ex}}\left(\mathbf{r}\right)$ and the magnitude of the field like the spin torque $\tau_{\text{f}}\left(\mathbf{r}\right)$
can be calculated analytically. The saturation magnetization can be expressed as 
\begin{equation}
	M_{s}\left(\boldsymbol{G}\right)=\begin{cases}
		\begin{array}{cc}
			f(M_{s,\text{Co}}-M_{s,\text{Py}})+M_{s,\text{Py}},\qquad & \text{for \ensuremath{G=0}}\\
			\\(M_{s,\text{Co}}-M_{s,\text{Py}})2f\cfrac{J_{1}\left(GR\right)}{GR} & \text{for \ensuremath{G\neq0}}
	\end{array}\end{cases}\label{eq:6}
\end{equation}

\begin{quote}
	where $J_{1}$ is a Bessel function of the first kind, $R$ is the
	radius of the dots, $M_{s,\text{Co}}\,\text{and}\,M_{s,\text{Py}}$
	are the saturation magnetization in Co and Py respectively. $f$ is
	the filling fraction defined as the ratio of the area occupied by
	Co dots to the area of the unit cell, $f=\pi r^{2}/a^{2}$. The formula
	for $\lambda^{2}_{\text{ex}}\left(\mathbf{G}\right)$ and $\tau_{\text{f}}\left(\mathbf{G}\right)$
	has the same form as Eq.(\ref{eq:6}). The static and dynamic magnetostatic
	fields can be derived in the form\cite{sokolovskyy2011magnetostatic,kaczer1974demagnetizing}
\begin{equation}
	\begin{aligned}
		H_{\mathrm{ms},z}
		(\mathbf{r}_{\parallel},x)
		=
		-\sum_{\mathbf{G}}
		&
		\frac{M_{s}(\mathbf{G})}{|\mathbf{G}|^{2}}
		G_{z}^{2}
		\left[
		1-
		\cosh
		\left(
		|\mathbf{G}|x
		\right)
		e^{-|\mathbf{G}|d/2}
		\right]
		\\
		&
		\times
		e^{i\mathbf{G}\cdot\mathbf{r}_{\parallel}}
		\label{eq:7}
	\end{aligned}
\end{equation}

\begin{equation}
	\begin{aligned}
		h_{\mathrm{ms},y}
		(\mathbf{r}_{\parallel},x)
		=
		-\sum_{\mathbf{G}}
		&
		\Bigg[
		\frac{
			m_{y}[\mathbf{G}]
		}{
			|\mathbf{k}+\mathbf{G}|^{2}
		}
		(k_{y}+G_{y})^{2}
		\\
		&
		\times
		\left(
		1-
		\cosh
		\left[
		|\mathbf{k}+\mathbf{G}|x
		\right]
		\right)
		e^{-|\mathbf{k}+\mathbf{G}|d/2}
		\Bigg]
		\\
		&
		\times
		e^{i(\mathbf{k}+\mathbf{G})\cdot\mathbf{r}_{\parallel}}
		\label{eq:8}
	\end{aligned}
\end{equation}

\begin{equation}
	\begin{aligned}
		h_{\mathrm{ms},x}
		(\mathbf{r}_{\parallel},x)
		=
		-\sum_{\mathbf{G}}
		&
		\Bigg[
		m_{x}[\mathbf{G}]
		\cosh
		\left(
		|\mathbf{k}+\mathbf{G}|x
		\right)
		\\
		&
		\times
		e^{-|\mathbf{k}+\mathbf{G}|d/2}
		\Bigg]
		e^{i(\mathbf{k}+\mathbf{G})\cdot\mathbf{r}_{\parallel}}
		\label{eq:9}
	\end{aligned}
\end{equation}
	
The substitution of Eq. (\ref{eq:4}) to Eq. (\ref{eq:9}) into Eq. (\ref{eq:1}) leads to eigenvalue problem with eigenvalues $i2\pi f/(\gamma\mu_{0}H_{0})$:
	\begin{equation}
		\hat{M}m_{k}=i\cfrac{2\pi f}{\gamma\mu_{0}H_{0}}m_{k},\label{eq:10}
	\end{equation}
where the eigenvector is \[
\begin{aligned}
	m_{k}^{T}
	=
	[
	&
	m_{x,k}(\mathbf{G}_{1}),
	\ldots,
	m_{x,k}(\mathbf{G}_{N}),
	\\
	&
	m_{y,k}(\mathbf{G}_{1}),
	\ldots,
	m_{y,k}(\mathbf{G}_{N})
	]
\end{aligned}
\] when a finite number $N$ of reciprocal lattice vectors is used in the Fourier series. The elements of the matrix $\hat{M}$are specified in the Appendix. 
\end{quote}

\subsection*{Magnonic band structure in a 2D MC}

\begin{figure}[!ht]
	\centering \includegraphics[width=1\columnwidth]{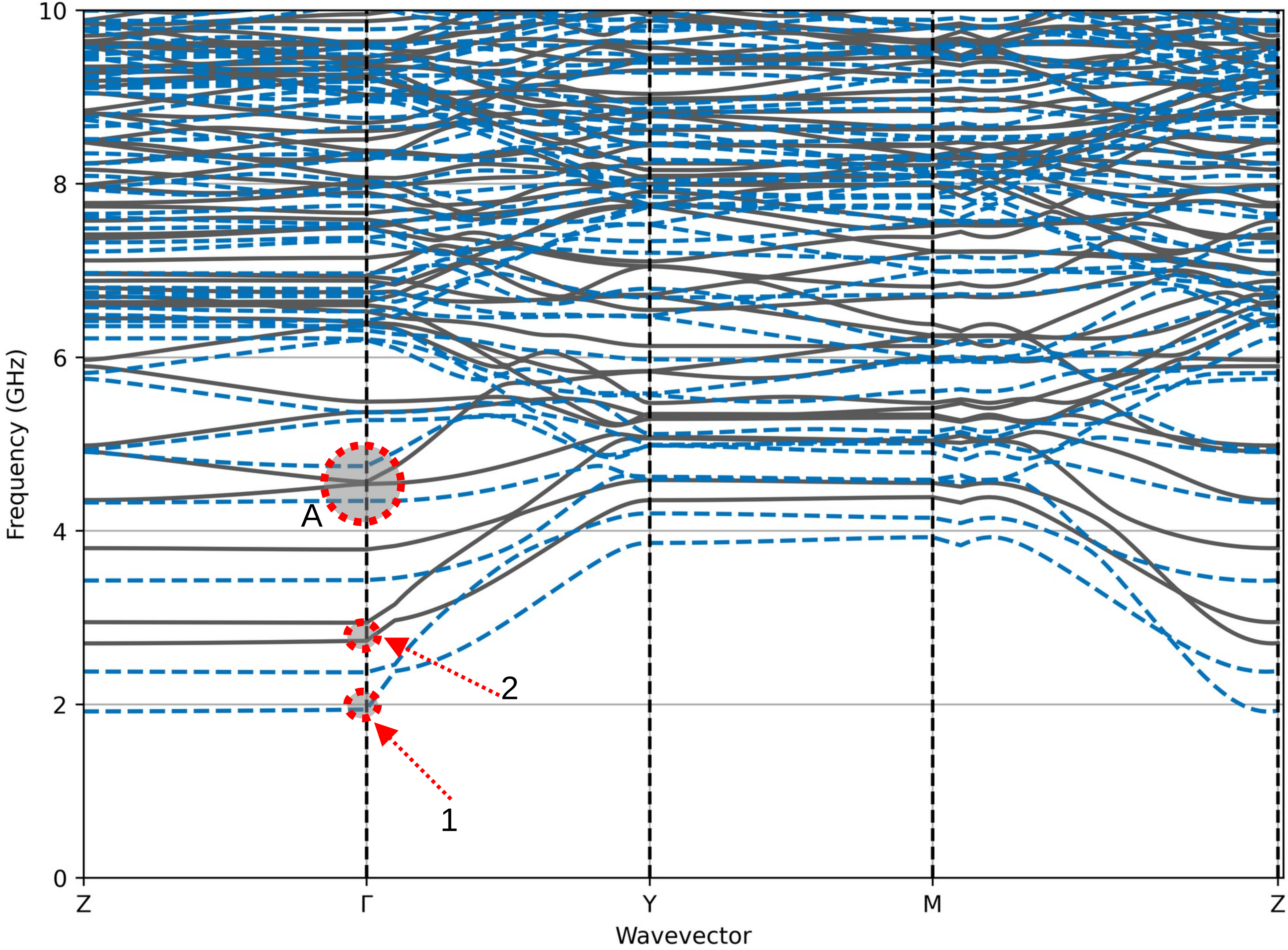} \caption{Magnonic band structure of 2D BMC with heavy metal layer interfaced with Py thin film calculated along the path in the first BZ shown in Fig. \ref{fig:(a)First-BZ-for}(a). The device is saturated along the $z$ axis by the static external magnetic field $\mu_{0}H_{0}=15\text{mT}$. The grey-coloured band structure is obtained without considering spin torque effects, and the blue-dotted band structure is obtained by incorporating the spin torque term in the PWM analysis. The capital letters A and B is the region of hybridization. The region A is enlarged in Fig. 4. The eigen mode profile for the point at 1 and 2 is depicted in Fig. 3.}\label{fig:Magnonic-band-structure}
\end{figure}

The magnonic structure calculated along the path $\text{Z}-\Gamma-Y-\text{M}-\text{Z}$
of the first BZ in Fig.\ref{fig:(a)First-BZ-for} (a) for the 2D BMC
composed of Co dots in Py is depicted in Fig. \ref{fig:Magnonic-band-structure}.

The saturation magnetizations and exchange constants in Py and Co is taken to the value used in \cite{tacchi2012forbidden}. $M_{s,\text{Py}}=0.78\times10^{6}\text{Am}^{-1}$,
$A_{Py}=1.3\times10^{-11}\text{J}\text{m}^{-1}$ and $M_{s,\text{Co}}=1.00\times10^{6}\text{Am}^{-1}$, $A_{Py}=2.0\times10^{-11}\text{J}\text{m}^{-1}$, respectively. The magnonic crystal is formed by a square array of Co dots with $\text{R=155 nm}$
fully embedded in a Py film as depicted in Fig. \ref{fig:(a)First-BZ-for}. The lattice constant and thickness of the magnonic crystal is taken
to be $a=600\text{nm}$ and $d=20\text{nm}$. The magnonic crystal is saturated in the in the plane of periodicity along the $z$ axis by an external magnetic field $\mu_{0}H_{0}=0.02\text{T}.$ The heavy-metal layer is set to a thickness of 5 nm,, and the field-like torque efficiency $c^{\perp}$ is assumed to be 0.28. The current density is taken to be $8\times10^{12}\text{A/m}^{2}$. Magnetocrystalline anisotropy for both magnetic materials is neglected through out the analysis.
The magnonic band shown in Fig. \ref{fig:Magnonic-band-structure}.
The spin wave dynamics in a two dimensional periodic magnonic crystal composed of Co nanodots embedded in a Py matrix by incorporating the
effect of spatially modulated field like torque term.

The spin-wave eigen modes in a two-dimensional periodic magnonic crystal composed of Co nanodots embedded in a Py matrix, incorporating the
effect of a spatially modulated field, such as a field like torque term, is extensively investigated. The magnonic band structure of
2D BMC along the high symmetry point is investigated using a plane-wave-based
eigenvalue approach\cite{Planewave_Maciej_2008}, as shown in Fig. \ref{fig:Magnonic-band-structure}. Without spin torque, the 2D BMC displays typical dispersive bands and only weak hybridisation. When
spin torque is applied, the dispersion changes noticeably. The field-like torque term acts as an effective magnetic field, normalising the magnonic
band structure and shifting. Strong couplings between localised and propagating spin-wave modes lead to several anticrossings. These anticrossings
create magnonic band gaps and cause strong hybridisation throughout the spectrum. Fig. \ref{fig:Magnonic-band-structure} reveals a clear
anti-crossing at $\Gamma$ point, indicating strong coupling between two magnonic modes with comparable frequencies(see the region marked
as A in Fig. \ref{fig:Magnonic-band-structure}). In the absence of interaction, these modes intersect. However periodic injection of
effective spin torque introduced by Co lattice, with the spin torque contributions, causes an effective coupling that leads to the band
level repulsion. As a results the band structure exhibit a finite frequency gap($\Delta f$) which directly reflect the strength of the interaction. In the low-frequency regime(4-5 GHz), the inclusion
of the spin torque term causes a mode-selective renormalisation of magnon dispersion. Periodic effective spin-torque injection leads
to mode-selective renormalisation of the magnon dispersion. The spin torque boosts inter-mode coupling at $\Gamma$ point, resulting in
increased band repulsion and the emergence of hybridization between adjacent modes. This appears as a measurable modification of the effective
anticrossing gap and a redistribution of spin wave modal character across the dispersion. The torque increases the anticrossing gap and induces strong hybridization, as evidenced by the exchange of modal character across the avoided crossing.

\subsection*{Spin torque induced tailoring of spin wave eigen modes}

The solution of Eq. \ref{eq:1} using plane wave method(PWM) yields both eigenfrequencies and eigenvectors $\mathbf{m}_{\mathbf{k}}$,
the latter being the Fourier coefficient of the dynamic magnetization distribution in 2D BMC. Spatial profiles of this components can be
determined on the basis of Bloch's theorem and the eigenvectors found as follows: 
\begin{equation}
	\mathbf{m}\left(\mathbf{r}\right)=\underset{\mathbf{G}}{\sum}m\left(\mathbf{G}\right)\text{e}^{-i\left(\mathbf{k}+\mathbf{G}\right)\cdot\mathbf{r}}\label{eq:11}
\end{equation}

\begin{figure}[!ht]
	\centering \includegraphics[width=1\columnwidth]{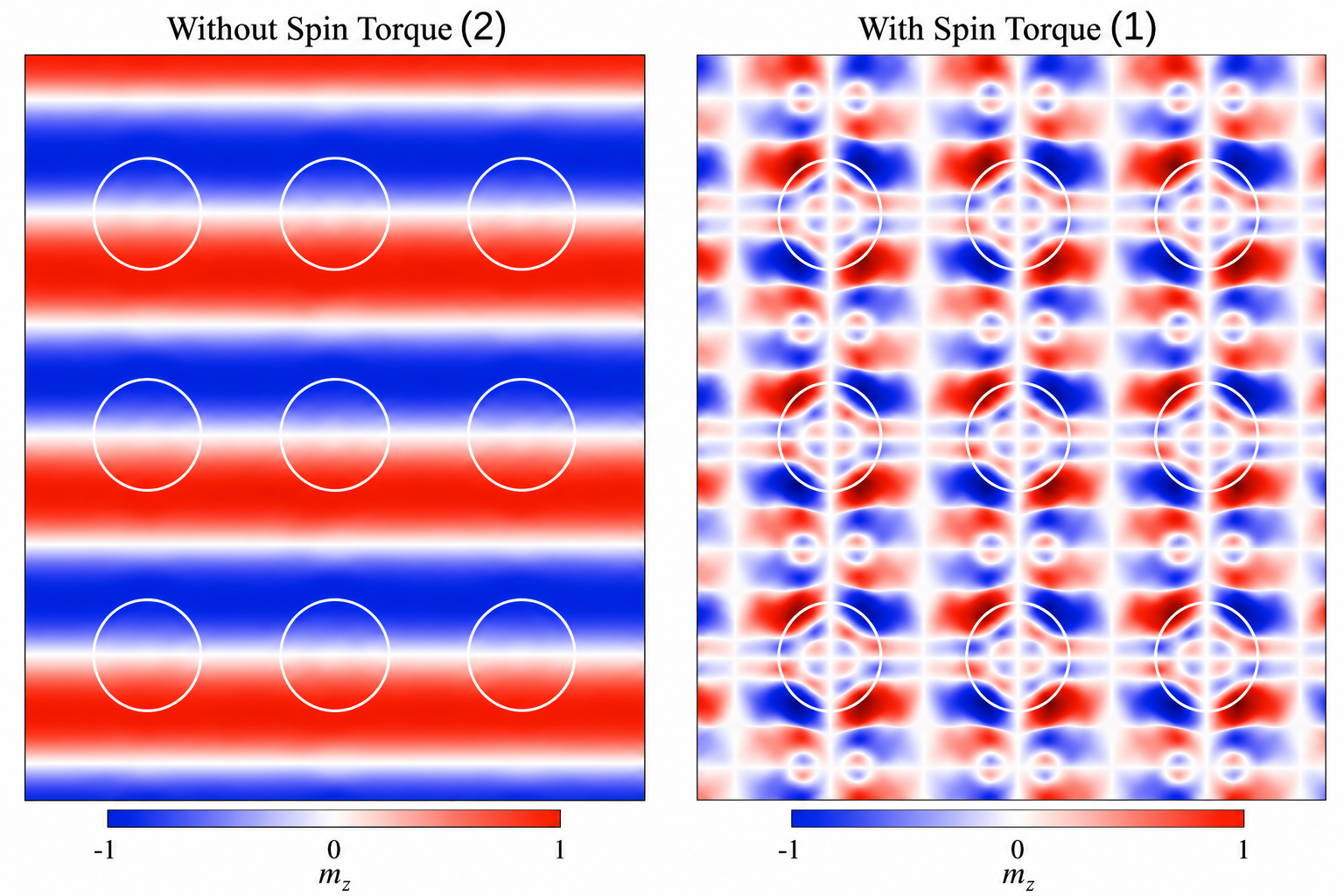}
	\caption{Magnon band resolved spin-wave eigenmodes profile corresponding to
		the selected points in the dispersion relation shown in Fig. \ref{fig:Magnonic-band-structure}(point
		1 and 2). }\label{fig:Magnon-band-resolved}
\end{figure}

To elucidate the magnetization distribution induced by the periodic injection of effective spin torque in 2D BMC, the spin-wave eigenmode
profiles were investigated across different regions of the band structure. To understand the microscopic origin of the spin torque driven band
renormalization of the low-frequency magnon bands generating from extra torque term in Landau-Lifshitz equation, the spatial magnetization
profiles associated with points 1 and 2 in Fig. \ref{fig:Magnonic-band-structure} at the $\Gamma$ point were systematically investigated. The amplitude of the dynamical component of the magnetisation is investigated using
coloured maps. In these maps, red indicates the highest spin-precessional amplitude and blue shows the lowest for the chosen wave vectors. The modes for which profiles are calculated are marked by bold circles on the dispersion curve in Fig. \ref{fig:Magnonic-band-structure}.
The spin-wave eigenmode profiles corresponding to the selected points in the dispersion relation shown in Fig. \ref{fig:Magnonic-band-structure}(point 1 and 2 at $\Gamma$) are presented in Fig. \ref{fig:Magnon-band-resolved}.

The eigenmode profile at point 2 in the dispersion (Fig. \ref{fig:Magnonic-band-structure}) without spin torque shows a delocalized, stripe-like standing wave pattern that extends across the magnonic crystals. In this case, the spin wave forms broad horizontal stripes, so the magnons move together through the structure with only minor disturbance from the periodic nanodots. The almost uniform spatial distribution suggests that the nanodot lattice acts as a weak scattering potential, leading to coherent Bloch modes with little mixing between higher-order harmonics in the
Fourier domain. When spin torque is applied, the situation changes a lot. Spin torque adds energy to the 2D BMC in a regular way, which increases how much different spin-wave components interact within the periodic lattice. As a result, the spin waves no longer exhibit the stripe-like pattern observed previously. Instead, the different
wave components interact more strongly, leading to a more complex spin wave profile, as shown in Fig. \ref{fig:Magnon-band-resolved}(right). This means the spin waves scatter more strongly from the periodic nanodot lattice, and different magnon modes start to couple. The magnetic distribution then becomes a mixed or hybrid collective state, not just a simple stripe pattern. The spin wave profile clearly shows stronger wave interference, more mode coupling, and greater interaction
with the periodic magnetic crystal lattice. Periodic injection of spin torque causes only a small change in magnon dispersion at the
lowest frequencies, but the related eigenmode profiles reveal significant changes in the spin wave pattern. This means that spin torque increases
the interaction between different spin-wave modes, leading to a major change in the magnonic eigenstates.

\subsection*{Spin torque induced magnon band repulsion at high symmetry point}

\begin{figure}[!ht]
	\centering \includegraphics[width=1\columnwidth]{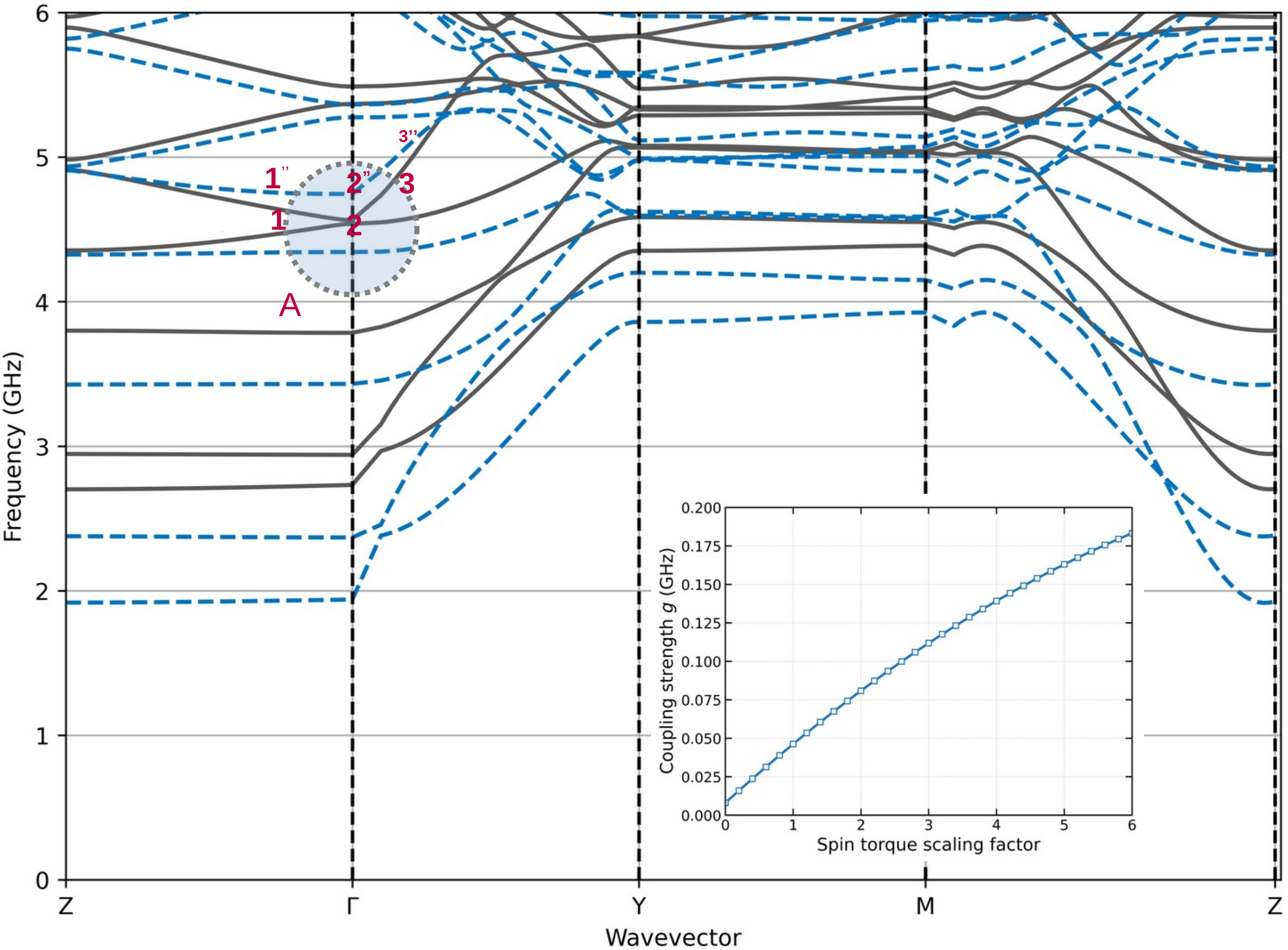}
	\caption{Magnonic band structures around the anti-crossings region, The zoomed
		version of band structure shown in Fig. \ref{fig:Magnonic-band-structure}.
		The coupling factor($g=\Delta f/2$) versus different spin torque
		scaling factor is shown as inset. The inset shows how the proposed
		spin-torque-induced magnonic crystal can be electrically tuned.}\label{fig:Magnonic-band-structures}
\end{figure}

This section discusses the magnonic band structures near the anti-crossing region. A detailed view of the band structure is shown in the Fig. \ref{fig:Magnonic-band-structures}

Points 1 and $1^{''}$ correspond to the magnon modes before hybridization in the absense and presence of spin torque, respectively. Similarly,
point 2 and $2^{''}$ represent the magnon modes at hybridization region in the absense and presence of spin torque. Points 3 and $3''$ indicate the position after hybridisation, while points 3 and $3^{''}$denote the modes after hybridization for the case without and with spin torque.
When two magnon modes come close in frequency, the periodic injection of effective spin torque allows these modes to interact and exchange energy opposing them from remaining independent. Instead of crossing, the system forms new hybrid mode in a stabilized manner that are combination of original ones. To stabilize these new states, their frequency split apart and forming band gap. The separation between these mode is denoted
as coupling strength $g$. The band gap increases with increase in spin torque parameter. The band structure exhibits a clear anti-crossing
feature at the $\Gamma$ point, showing strong coupling between two magnon modes with nearby frequencies. This mainly occurs through the
energy-exchange mechanism, driven by the periodic injection of effective spin torque into the magnonic lattice. In the absence of interaction,
these two modes intersect and retain their original mode character. Since the coupling strength between the two modes at the crossing
point is very weak, there is no energy exchange, and they retain their original mode character. This happens at the $\Gamma$ point between
4-5 GHz when there is no periodic injection of effective spin torque into the magnonic lattice.
Figure \ref{fig:Magnonic-band-structures} (inset) shows the coupling strength for different injection current densities, given as the spin
torque scaling factor. The hybridisation gap ($g=\Delta f/2$) extracted at $\Gamma$ point increases monotonically with increasing spin torque
strength. These results show that spin torque increases the coupling between interacting magnon branches. The graph demonstrates that the
magnon coupling strength $g$ can be adjusted using spin torque. This tunability enables the creation of electrically reconfigurable magnonic-crystal
devices for adaptive microwave signal processing and programmable spin-wave computing. Controlling magnon hybridisation strength with periodic effective spin-torque injection also enables coherent manipulation of coupled magnon states in hybrid magnonic and quantum information
systems. 

\subsection*{Spin torque induced hybridization of magnon eigen modes in a magnonic
	lattice}

\begin{figure}[!ht]

 \includegraphics[width=1\columnwidth]{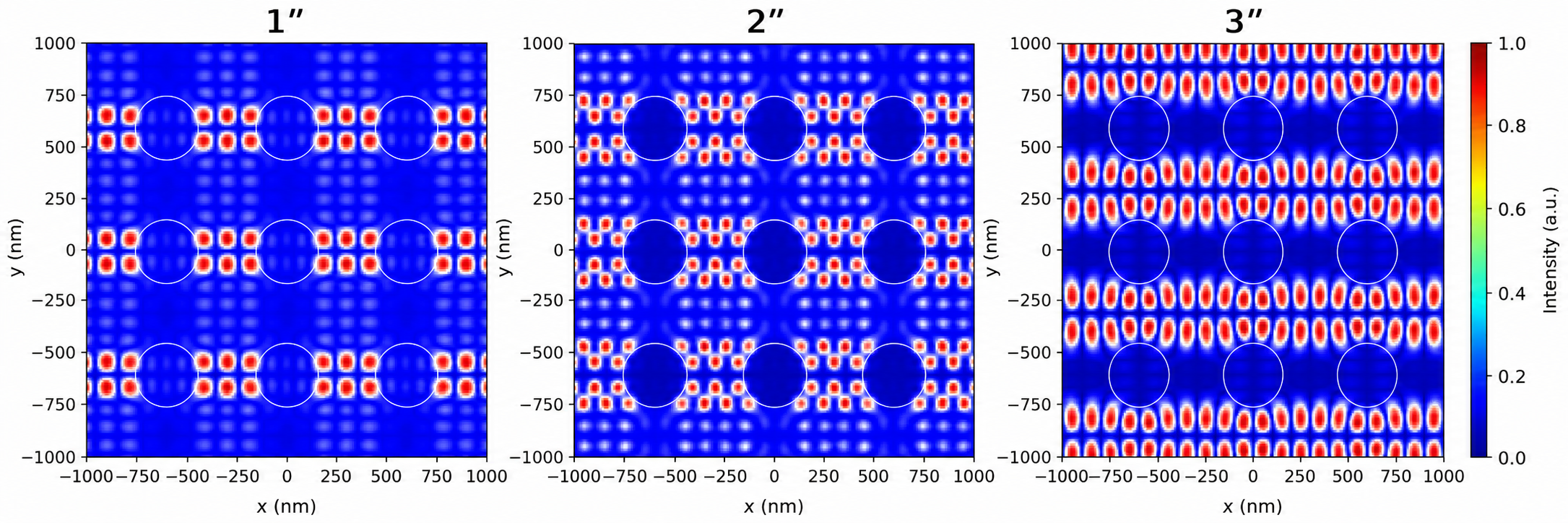}\caption{Spin wave eigen mod evolution at different point in the dispersion
 in Fig. \ref{fig:Magnonic-band-structures} under periodic injection of effective spin torque term into magnonic lattice. The white circles show the regular pattern of Co dots placed in Py. The red and blue areas show how the phase and amplitude of the dynamic magnetisation are distributed.}\label{fig:Spin-wave-eigen}
\end{figure}

Fig. \ref{fig:Spin-wave-eigen} shows how spin wave eigen modes change in different regions of the dispersion, as marked at region A in Fig.
\ref{fig:Spin-wave-eigen}. It tracks the evolution of these modes as the magnonic crystal moves from the prehybridised to the hybridised regime, and then to the post-hybridisation region in the magnonic lattice, all under periodic effective spin-torque excitation.

The spatial magnetization distributions are shown for three points chosen from the dispersion relation. These points are labelled as before-hybridisation ($1^{''}$), hybridised ($2^{''}$), and after-hybridisation ($3^{''}$) regions. In the pre-hybridised region ($1^{''}$), the eigenmode exhibits a simple stripe-like pattern, mainly along the vertical direction. The spin wave patterns are regular between neighbouring
nanodots. This tells us that spin waves are propagating with very weak interaction between neighbouring modes. In this situation, modes
retain their original shape and show only a small amount of wave mixing occurring at this stage. At the hybridised stage, the spin wave mode
shows a clear change in its spatial distribution compared to the pre-hybridised state. The vertical stripe, which we saw in the prehybridized state,
becomes distorted, and the spin-wave profile spreads over a larger region. This indicates that nearby magnon bands begin to interact strongly at this point. As a result, the mode shape changes compared to the pre-hybridised state, and the spin wave begins to exhibit characteristics
of interacting modes. At point $3^{''}$, the spin wave mode develops a different spatial distribution compared to the pre-hybridised and
hybridised states. The magnetisation direction becomes more directional in the magnetic inclusions. This clearly shows that the original mode structure is modified after hybridisation when it interacts with magnon
modes. As a result, the spin wave mode no longer maintains its initial state and transitions to a new propagating state due to the spin torque
effect. As magnetic lattices change from a prehybridised to a post-hybridised state, the spin wave modes shift from being localised to spreading
out and propagating. Before hybridisation, spin waves remained confined to a small region around the nanodots. After hybridisation, the eigenmodes
stretch further along the direction of propagation, indicating stronger spin-wave motion driven by spin-torque effects. The result indicate
that propagation characteristics of spin wave can be easily controlled using injection of spin torque into magnetic lattice in a periodic
manner. The transformation from relatively localized states to a more propagating modes tells us that the spin wave transport properties
inside the magnetic lattice can be easily controlled through electrical
spin injection.

\begin{figure}[!ht]
	\centering \includegraphics[width=1\columnwidth]{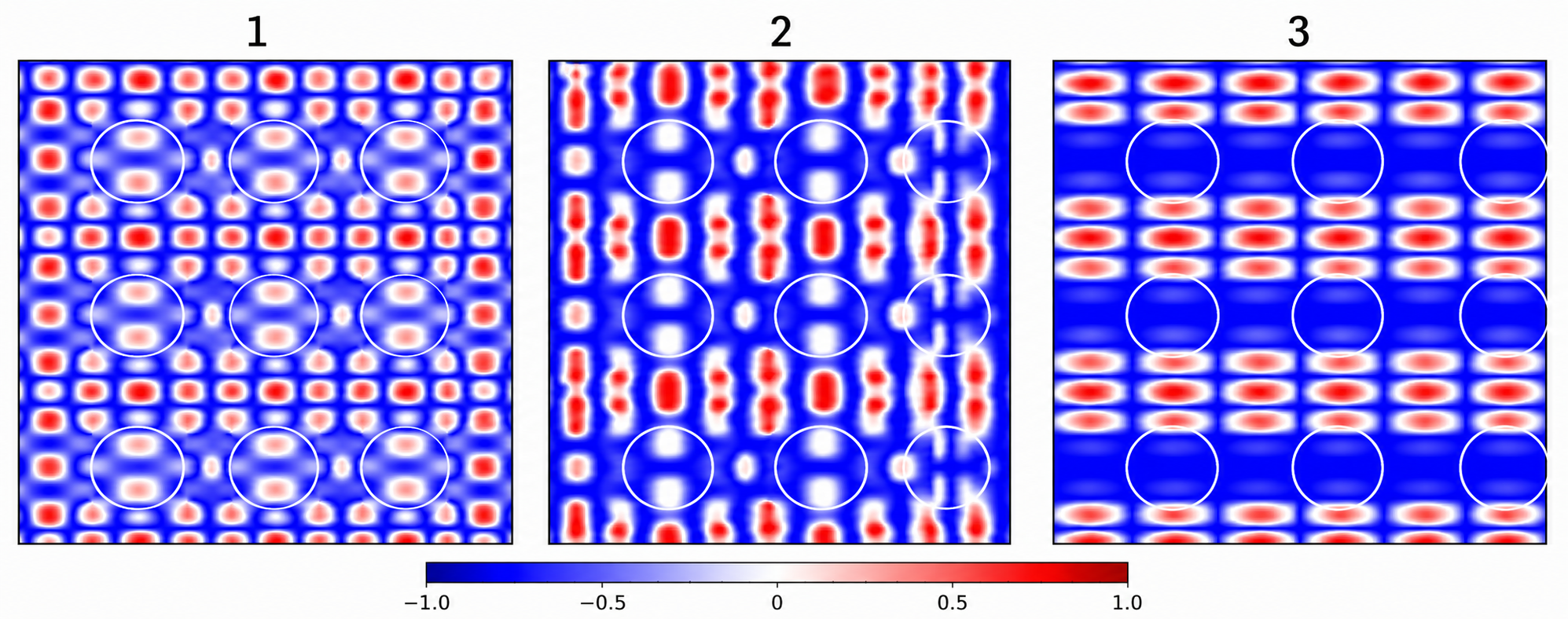}\caption{Spin wave eigen mod evolution at different point in the dispersion in Fig. \ref{fig:Magnonic-band-structures} without spin torque effect into magnonic lattice. The white circles show the regular pattern
		of Co dots placed in Py. The red and blue areas show how the phase and amplitude of the dynamic magnetisation are distributed.}\label{fig:Spin-wave-eigen-without-spin-torque}
\end{figure}

Fig. \ref{fig:Spin-wave-eigen-without-spin-torque} represents the evolution of spin wave modes without spin torque term in the magnonic lattice.

At point 1, the eigenmode maintains its original propagating characteristics, exhibiting only weak interaction with adjacent magnon bands. At point 2, the two magnon bands converge in frequency at the $\Gamma$ point. The mode profile undergoes minimal change compared to the pre-crossing
state, which suggests a weak interaction between neighboring bands. The spin wave continues to propagate as previously observed, indicating
that the coupling between bands remains weak in the absence of periodic effective spin-torque injection within the lattice. This crossing
primarily demonstrates weak mode interaction rather than strong hybridisation. At point 3, the spin wave mode persists with a spatial profile closely resembling the pre-crossing state, although a slight modification in intensity distribution is observed after the crossing region. The overall propagating nature of the spin wave modes is preserved. These observations indicate that the interacting bands largely retain their original mode character, confirming that the crossing occurs with only weak coupling when spin torque is absent.

\section*{Summary}

In summary, electrically tunable mode coupling and band hybridisation in a bicomponent magnonic crystal via inhomogeneous current-induced
spin torque are demonstrated using the plane-wave method. A substantial modification of the magnonic band structure was observed, including
tunable anticrossing, band deformation, and a spin-torque-induced hybridisation gap. The periodic injection of effective spin torque
into magnonic lattices enhances the dynamic manipulation of the interaction between localised and propagating Damon-Esbach modes, resulting in
enhanced mode mixing and electrically reconfigurable magnonic states. The band structure exhibits good tunablity with an external current
induced inhomogeneous spin torque injection, causes an effective coupling that leads to magnonic band repulsion. It is observed that coupling strength monotonically increasing with an increase in spin torque scaling factor. This study closely examines how the eigenmodes in a magnonic lattice change as they move through the hybridisation region. The results show that when the lattice is excited by periodic effective
spin-torque, a confined lattice mode starts to behave more like a propagating mode. Both the position and size of the anticrossing gap
changed when an external current was applied, which offers a clear way to actively engineer magnonic bands. These findings show that
inhomogeneous spin torque is an effective way to control nanoscale spin-wave movement and could help create reconfigurable magnonic devices,
adaptive nanoscale microwave parts, and wave-based information processing technologies. The results show that spin torque can effectively control
magnonic band engineering using electrical signals. This method may help advance quantum magnonics and coherent wave-based computing systems.

\section*{Acknowledgements}

The simulations were carried out using the Madhava High Performance Computing (HPC) cluster at the Centre for Computational Modelling and Simulation (CCMS), National Institute of Technology Calicut, India.

\section*{Appendix: Matrix elements of the eigenvalue problem}

The matrix $\hat{M}$ of the eigenvalue problem problem (\ref{eq:9})
can be written in a block matrix form: 
\begin{equation}
	\hat{M}=\left(\begin{array}{cc}
		\hat{M}^{xx} & \hat{M}^{xy}\\
		\hat{M}^{yx} & \hat{M}^{yy}
	\end{array}\right)\label{eq:12}
\end{equation}
The submatrices in (\ref{eq:12}) are defined as follows:, 
\begin{equation}
	\hat{M}^{xx}_{ij}=\hat{M}^{yy}_{ij}\label{eq:13}
\end{equation}

\begin{equation}
	\begin{aligned}
		M_{ij}^{xy}
		=
		&
		\delta_{ij}
		+
		\sum_{l}
		\frac{
			(\mathbf{k}+\mathbf{G}_{j})
			\cdot
			(\mathbf{k}+\mathbf{G}_{l})
		}{
			H_{0}
		}
		\lambda_{\mathrm{ex}}^{2}
		(\mathbf{G}_{l}-\mathbf{G}_{j})
		M_{s}
		(\mathbf{G}_{i}-\mathbf{G}_{l})
		\\
		&
		+
		\frac{
			(k_{y}+G_{y,j})^{2}
		}{
			H_{0}
			|\mathbf{k}+\mathbf{G}_{j}|^{2}
		}
		\left[
		1-
		C(\mathbf{k}+\mathbf{G}_{j},x)
		\right]
		M_{s}
		(\mathbf{G}_{i}-\mathbf{G}_{j})
		\\
		&
		+
		\tau_{f}
		(\mathbf{G}_{i}-\mathbf{G}_{j})
		\\
		&
		-
		\frac{
			(G_{z,i}-G_{z,j})
		}{
			|\mathbf{G}_{i}-\mathbf{G}_{j}|^{2}
		}
		M_{s}
		(\mathbf{G}_{i}-\mathbf{G}_{j})
		\\
		&
		\times
		\left[
		1-
		C(\mathbf{G}_{i}-\mathbf{G}_{j},x)
		\right]
		\label{eq:14}
	\end{aligned}
\end{equation}

\begin{equation}
	\begin{aligned}
		M_{ij}^{yx}
		=
		&
		-\delta_{ij}
		-
		\sum_{l}
		\frac{
			(\mathbf{k}+\mathbf{G}_{j})
			\cdot
			(\mathbf{k}+\mathbf{G}_{l})
		}{
			H_{0}
		}
		\lambda_{\mathrm{ex}}^{2}
		(\mathbf{G}_{l}-\mathbf{G}_{j})
		M_{s}
		(\mathbf{G}_{i}-\mathbf{G}_{l})
		\\
		&
		-
		\frac{
			1
		}{
			H_{0}
		}
		C(\mathbf{k}+\mathbf{G}_{j},x)
		M_{s}
		(\mathbf{G}_{i}-\mathbf{G}_{j})
		\\
		&
		+
		\frac{
			(G_{z,i}-G_{z,j})
		}{
			|\mathbf{G}_{i}-\mathbf{G}_{j}|^{2}
		}
		M_{s}
		(\mathbf{G}_{i}-\mathbf{G}_{j})
		\\
		&
		\times
		\left[
		1-
		C(\mathbf{G}_{i}-\mathbf{G}_{j},x)
		\right]
		\label{eq:15}
	\end{aligned}
\end{equation}

\noindent where indexes of reciprocal lattice vectors $i,j,l$ are integer numbers. The additional function used in the equations above
is defined as follows: 
\begin{equation}
	C\left(k,x\right)=\cosh\left(|\boldsymbol{k}|x\right)e^{-|k|d/2}\label{eq:16}
\end{equation}

\bibliography{paper}

@book{joannopoulos2008,
  author    = {Joannopoulos, J. D. and Meade, R. D. and Winn, J. N.},
  title     = {Photonic Crystals: Molding the Flow of Light},
  edition   = {2},
  publisher = {Princeton University Press},
  address   = {Princeton, NJ},
  year      = {2008}
}

@book{prather2009,
  author    = {Prather, D. W. and Shi, S. and Sharkawy, A. and Murakowski, J. and Schneider, G. J.},
  title     = {Photonic Crystals: Theory, Applications, and Fabrication},
  publisher = {Wiley},
  address   = {New York},
  year      = {2009}
}

@article{krawczyk2008,
  author  = {Krawczyk, M. and Puszkarski, H.},
  journal = {Phys. Rev. B},
  volume  = {77},
  pages   = {054437},
  year    = {2008}
}

@article{tiwari2010,
  author  = {Tiwari, R. P. and Stroud, D.},
  journal = {Phys. Rev. B},
  volume  = {81},
  pages   = {220403},
  year    = {2010},
  note    = {(R)}
}

@article{klos2012,
  author  = {Klos, J. W. and Sokolovskyy, M. L. and Mamica, S. and Krawczyk, M.},
  journal = {J. Appl. Phys.},
  volume  = {111},
  pages   = {123910},
  year    = {2012}
}

@article{mamica2012,
  author  = {Mamica, S. and Krawczyk, M. and Sokolovskyy, M. L. and Romero-Vivas, J.},
  journal = {Phys. Rev. B},
  volume  = {86},
  pages   = {144402},
  year    = {2012}
}

@article{yang2012,
  author  = {Yang, H. and Yun, G. and Cao, Y.},
  journal = {J. Appl. Phys.},
  volume  = {111},
  pages   = {013908},
  year    = {2012}
}

@book{ashcroft1976,
  author    = {Ashcroft, N. W. and Mermin, N. D.},
  title     = {Solid State Physics},
  publisher = {Holt, Rinehart and Winston},
  address   = {New York},
  year      = {1976}
}

@book{yariv2003,
  author    = {Yariv, A. and Yeh, P.},
  title     = {Optical Waves in Crystals},
  publisher = {Wiley-Interscience},
  address   = {New York},
  year      = {2003},
  note      = {p. 156}
}

@article{kichin2012,
  author  = {Kichin, G. and Weiss, T. and Gao, H. and Henzie, J. and Odom, T. W. and Tikhodeev, S. G. and Giessen, H.},
  journal = {Physica B},
  volume  = {407},
  pages   = {4037},
  year    = {2012}
}

@article{tartakovskaya2000,
  author  = {Tartakovskaya, E.},
  journal = {Phys. Rev. B},
  volume  = {62},
  pages   = {11225},
  year    = {2000}
}

@article{dyakonov1971current,
  title={Current-induced spin orientation of electrons in semiconductors},
  author={Dyakonov, Mikhail I and Perel, VI},
  journal={Physics Letters A},
  volume={35},
  number={6},
  pages={459--460},
  year={1971},
  publisher={Elsevier}
}

@article{jungwirth2012spin,
  title={Spin Hall effect devices},
  author={Jungwirth, Tomas and Wunderlich, J{\"o}rg and Olejn{\'\i}k, Kamil},
  journal={Nature materials},
  volume={11},
  number={5},
  pages={382--390},
  year={2012},
  publisher={Nature Publishing Group UK London}
}

@article{murakami2003dissipationless,
  title={Dissipationless quantum spin current at room temperature},
  author={Murakami, Shuichi and Nagaosa, Naoto and Zhang, Shou-Cheng},
  journal={Science},
  volume={301},
  number={5638},
  pages={1348--1351},
  year={2003},
  publisher={American Association for the Advancement of Science}
}

@article{valenzuela2006direct,
  title={Direct electronic measurement of the spin Hall effect},
  author={Valenzuela, Sergio O and Tinkham, Michael},
  journal={Nature},
  volume={442},
  number={7099},
  pages={176--179},
  year={2006},
  publisher={Nature Publishing Group UK London}
}

@article{krawczyk2012formulation,
  title={On the Formulation of the Exchange Field in the Landau-Lifshitz Equation for Spin-Wave Calculation in Magnonic Crystals},
  author={Krawczyk, Maciej and Sokolovskyy, ML and Klos, JW and Mamica, S{\l}awomir},
  journal={Advances in Condensed Matter Physics},
  volume={2012},
  number={1},
  pages={764783},
  year={2012},
  publisher={Wiley Online Library}
}

@article{sokolovskyy2011magnetostatic,
  title={The magnetostatic modes in planar one-dimensional magnonic crystals with nanoscale sizes},
  author={Sokolovskyy, ML and Krawczyk, Maciej},
  journal={Journal of Nanoparticle Research},
  volume={13},
  number={11},
  pages={6085--6091},
  year={2011},
  publisher={Springer}
}

@article{kaczer1974demagnetizing,
  title={On the demagnetizing energy of periodic magnetic distributions},
  author={Kaczer, J and Murtinova, L},
  journal={physica status solidi (a)},
  volume={23},
  number={1},
  pages={79--86},
  year={1974},
  publisher={Wiley Online Library}
}

@article{tacchi2012forbidden,
  title={Forbidden Band Gaps in the Spin-Wave Spectrum of a Two-Dimensional Bicomponent Magnonic Crystal},
  author={Tacchi, Silvia and Duerr, G and Klos, JW and Madami, Marco and Neusser, S and Gubbiotti, Gianluca and Carlotti, Giovanni and Krawczyk, Maciej and Grundler, D},
  journal={Physical review letters},
  volume={109},
  number={13},
  pages={137202},
  year={2012},
  publisher={APS}
}

@article{Planewave_Maciej_2008,
  title = {Plane-wave theory of three-dimensional magnonic crystals},
  author = {Krawczyk, M. and Puszkarski, H.},
  journal = {Phys. Rev. B},
  volume = {77},
  issue = {5},
  pages = {054437},
  numpages = {13},
  year = {2008},
  month = {Feb},
  publisher = {American Physical Society},
}

@article{Intro_1_kruglyak2010magnonics,
  title={Magnonics},
  author={Kruglyak, VV and Demokritov, SO and Grundler, D},
  journal={Journal of Physics D: Applied Physics},
  volume={43},
  number={26},
  pages={264001},
  year={2010}
}

@article{intro_2_lenk2011building,
  title={The building blocks of magnonics},
  author={Lenk, Benjamin and Ulrichs, Henning and Garbs, Fabian and M{\"u}nzenberg, Markus},
  journal={Physics Reports},
  volume={507},
  number={4-5},
  pages={107--136},
  year={2011},
  publisher={Elsevier}
}

@article{Intro_3_chumak2015magnon,
  title={Magnon spintronics},
  author={Chumak, Andrii V and Vasyuchka, Vitaliy I and Serga, Alexander A and Hillebrands, Burkard},
  journal={Nature physics},
  volume={11},
  number={6},
  pages={453--461},
  year={2015},
  publisher={Nature Publishing Group UK London}
}

@book{Intro_4_demokritov2012magnonics,
  title={Magnonics: From fundamentals to applications},
  author={Demokritov, Sergej O and Slavin, Andrei N},
  volume={125},
  year={2012},
  publisher={Springer Science \& Business Media}
}

@article{Intro_5_khitun2011non,
  title={Non-volatile magnonic logic circuits engineering},
  author={Khitun, Alexander and Wang, Kang L},
  journal={Journal of Applied Physics},
  volume={110},
  number={3},
  year={2011},
  publisher={AIP Publishing}
}

@article{Intro_6_chumak2017magnonic,
  title={Magnonic crystals for data processing},
  author={Chumak, Andrii V and Serga, Alexander A and Hillebrands, Burkard},
  journal={Journal of Physics D: Applied Physics},
  volume={50},
  number={24},
  pages={244001},
  year={2017},
  publisher={IOP Publishing}
}

@article{Intro_7_nikitov2001spin,
  title={Spin waves in periodic magnetic structures magnonic crystals},
  author={Nikitov, SA and Tailhades, Ph and Tsai, CS},
  journal={Journal of Magnetism and Magnetic Materials},
  volume={236},
  number={3},
  pages={320--330},
  year={2001},
  publisher={Elsevier}
}

@article{Intro_8_bauer2012spin,
  title={Spin caloritronics},
  author={Bauer, Gerrit EW and Saitoh, Eiji and Van Wees, Bart J},
  journal={Nature materials},
  volume={11},
  number={5},
  pages={391--399},
  year={2012},
  publisher={Nature Publishing Group UK London}
}

@article{Intro_9_krawczyk2014review,
  title={Review and prospects of magnonic crystals and devices with reprogrammable band structure},
  author={Krawczyk, Maciej and Grundler, D},
  journal={Journal of physics: Condensed matter},
  volume={26},
  number={12},
  pages={123202},
  year={2014},
  publisher={IOP Publishing}
}

@article{Intro_10_grundler2015reconfigurable,
  title={Reconfigurable magnonics heats up},
  author={Grundler, Dirk},
  journal={Nature Physics},
  volume={11},
  number={6},
  pages={438--441},
  year={2015},
  publisher={Nature Publishing Group UK London}
}

@article{Intro_11_mruczkiewicz2017spin,
  title={Spin-wave nonreciprocity and magnonic band structure in a thin permalloy film induced by dynamical coupling with an array of Ni stripes},
  author={Mruczkiewicz, Micha{\l} and Graczyk, Piotr and Lupo, P and Adeyeye, A and Gubbiotti, G and Krawczyk, M},
  journal={Physical Review B},
  volume={96},
  number={10},
  pages={104411},
  year={2017},
  publisher={APS}
}

@article{Intro_12_gruszecki2015influence,
  title={Influence of magnetic surface anisotropy on spin wave reflection from the edge of ferromagnetic film},
  author={Gruszecki, Pawe{\l} and Dadoenkova, Yu S and Dadoenkova, NN and Lyubchanskii, IL and Romero-Vivas, Javier and Guslienko, KY and Krawczyk, Maciej},
  journal={Physical Review B},
  volume={92},
  number={5},
  pages={054427},
  year={2015},
  publisher={APS}
}

@article{Intro_13_krawczyk2008plane,
  title={Plane-wave theory of three-dimensional magnonic crystals},
  author={Krawczyk, M and Puszkarski, H},
  journal={Physical Review B Condensed Matter and Materials Physics},
  volume={77},
  number={5},
  pages={054437},
  year={2008},
  publisher={APS}
}

@article{Intro_14_klos2013magnonic,
  title={Magnonic band engineering by intrinsic and extrinsic mirror symmetry breaking in antidot spin-wave waveguides},
  author={K{\l}os, Jaros{\l}aw W and Kumar, Dheeraj and Krawczyk, Maciej and Barman, Anjan},
  journal={Scientific Reports},
  volume={3},
  number={1},
  pages={2444},
  year={2013},
  publisher={Nature Publishing Group UK London}
}

@article{Intro_15_rychly2015magnonic,
  title={Magnonic crystals prospective structures for shaping spin waves in nanoscale},
  author={Rych{\l}y, Justyna and Gruszecki, Pawe{\l} and Mruczkiewicz, Micha{\l} and K{\l}os, Jaros{\l}aw W and Mamica, S{\l}awomir and Krawczyk, Maciej},
  journal={Low Temperature Physics},
  volume={41},
  number={10},
  pages={745--759},
  year={2015},
  publisher={AIP Publishing}
}

@article{Intro_16_wagner2016magnetic,
  title={Magnetic domain walls as reconfigurable spin-wave nanochannels},
  author={Wagner, Kai and K{\'a}kay, A and Schultheiss, K and Henschke, A and Sebastian, T and Schultheiss, H},
  journal={Nature nanotechnology},
  volume={11},
  number={5},
  pages={432--436},
  year={2016},
  publisher={Nature Publishing Group UK London}
}

@article{Intro_17_sadovnikov2015magnonic,
  title={Magnonic beam splitter: The building block of parallel magnonic circuitry},
  author={Sadovnikov, Alexandr V and Davies, Carl S and Grishin, Sergey V and Kruglyak, VV and Romanenko, DV and Sharaevskii, Yu P and Nikitov, SA},
  journal={Applied Physics Letters},
  volume={106},
  number={19},
  year={2015},
  publisher={AIP Publishing}
}

@article{Intro_18_tacchi2012forbidden,
  title={Forbidden Band Gaps in the Spin-Wave Spectrum of a Two-Dimensional<? format?> Bicomponent Magnonic Crystal},
  author={Tacchi, Silvia and Duerr, G and Klos, JW and Madami, Marco and Neusser, S and Gubbiotti, Gianluca and Carlotti, Giovanni and Krawczyk, Maciej and Grundler, D},
  journal={Physical review letters},
  volume={109},
  number={13},
  pages={137202},
  year={2012},
  publisher={APS}
}

@article{Intro_19_neusser2009magnonics,
  title={Magnonics: Spin waves on the nanoscale},
  author={Neusser, Sebastian and Grundler, Dirk},
  journal={Advanced materials},
  volume={21},
  number={28},
  pages={2927--2932},
  year={2009},
  publisher={Wiley Online Library}
}

@article{Intro_20_mruczkiewicz2016collective,
  title={Collective dynamical skyrmion excitations in a magnonic crystal},
  author={Mruczkiewicz, Micha{\l} and Gruszecki, Pawe{\l} and Zelent, Mateusz and Krawczyk, Maciej},
  journal={Physical Review B},
  volume={93},
  number={17},
  pages={174429},
  year={2016},
  publisher={APS}
}

@article{Intro_21_ciubotaru2016all,
  title={All electrical propagating spin wave spectroscopy with broadband wavevector capability},
  author={Ciubotaru, Florin and Devolder, Thibaut and Manfrini, Mauricio and Adelmann, Christoph and Radu, IP},
  journal={Applied Physics Letters},
  volume={109},
  number={1},
  year={2016},
  publisher={AIP Publishing}
}

@article{Intro_22_khitun2010magnonic,
  title={Magnonic logic circuits},
  author={Khitun, Alexander and Bao, Mingqiang and Wang, Kang L},
  journal={Journal of Physics D: Applied Physics},
  volume={43},
  number={26},
  pages={264005},
  year={2010}
}

@article{Intro_23_chumak2014magnon,
  title={Magnon transistor for all-magnon data processing},
  author={Chumak, Andrii V and Serga, Alexander A and Hillebrands, Burkard},
  journal={Nature communications},
  volume={5},
  number={1},
  pages={4700},
  year={2014},
  publisher={Nature Publishing Group UK London}
}

@article{intro_24_schneider2008realization,
  title={Realization of spin-wave logic gates},
  author={Schneider, Thomas and Serga, Alexander A and Leven, Britta and Hillebrands, Burkard and Stamps, Robert L and Kostylev, Mikhail P},
  journal={Applied Physics Letters},
  volume={92},
  number={2},
  year={2008},
  publisher={AIP Publishing}
}

@article{intro_25_fulara2019spin,
  title={Spin-orbit torque--driven propagating spin waves},
  author={Fulara, Himanshu and Zahedinejad, Mohammad and Khymyn, Roman and Awad, AA and Muralidhar, Shreyas and Dvornik, Mykola and {\AA}kerman, Johan},
  journal={Science advances},
  volume={5},
  number={9},
  pages={eaax8467},
  year={2019},
  publisher={American Association for the Advancement of Science}
}

@article{Intro_26_khitun2005nano,
  title={Nano scale computational architectures with spin wave bus},
  author={Khitun, Alexander and Wang, Kang L},
  journal={Superlattices and Microstructures},
  volume={38},
  number={3},
  pages={184--200},
  year={2005},
  publisher={Elsevier}
}

@article{Intro_27_klingler2014design,
  title={Design of a spin-wave majority gate employing mode selection},
  author={Klingler, Stefan and Pirro, Philipp and Br{\"a}cher, Thomas and Leven, Britta and Hillebrands, Burkard and Chumak, Andrii V},
  journal={Applied Physics Letters},
  volume={105},
  number={15},
  year={2014},
  publisher={AIP Publishing}
}

@article{Intro_28_chumak2022advances,
  title={Advances in magnetics roadmap on spin-wave computing},
  author={Chumak, Andrii V and Kabos, Pavel and Wu, Mingzhong and Abert, Claas and Adelmann, Christoph and Adeyeye, Adekunle Olusola and {\AA}kerman, Johan and Aliev, Farkhad G and Anane, Abdelmadjid and Awad, Ahmad and others},
  journal={IEEE Transactions on Magnetics},
  volume={58},
  number={6},
  pages={1--72},
  year={2022},
  publisher={IEEE}
}

@article{Intro_29_grundler2016nanomagnonics,
  title={Nanomagnonics around the corner},
  author={Grundler, Dirk},
  journal={Nature nanotechnology},
  volume={11},
  number={5},
  pages={407--408},
  year={2016},
  publisher={Nature Publishing Group UK London}
}

@article{Intro_30_miron2011perpendicular,
  title={Perpendicular switching of a single ferromagnetic layer induced by in-plane current injection},
  author={Miron, Ioan Mihai and Garello, Kevin and Gaudin, Gilles and Zermatten, Pierre-Jean and Costache, Marius V and Auffret, St{\'e}phane and Bandiera, S{\'e}bastien and Rodmacq, Bernard and Schuhl, Alain and Gambardella, Pietro},
  journal={Nature},
  volume={476},
  number={7359},
  pages={189--193},
  year={2011},
  publisher={Nature Publishing Group UK London}
}

@article{Intro_31_liu2012spin,
  author={Liu, L. and others},
  title={Spin-Torque Switching with the Giant Spin Hall Effect},
  journal={Science},
  volume={336},
  pages={555--558},
  year={2012}
}

@article{Intro_32_hoffmann2013spin,
  author={Hoffmann, A.},
  title={Spin Hall Effects in Metals},
  journal={IEEE Transactions on Magnetics},
  volume={49},
  pages={5172--5193},
  year={2013}
}

@article{Intro_33_tserkovnyak2005spin,
  author={Tserkovnyak, Y. and others},
  title={Spin Pumping and Magnetization Dynamics},
  journal={Reviews of Modern Physics},
  volume={77},
  pages={1375--1421},
  year={2005}
}

@article{Intro_34_sinova2015spin,
  author={Sinova, J. and others},
  title={Spin Hall Effects},
  journal={Reviews of Modern Physics},
  volume={87},
  pages={1213--1260},
  year={2015}
}

@article{Intro_35_manchon2019current,
  author={Manchon, A. and others},
  title={Current-Induced Spin-Orbit Torques in ferromagnetic and antiferromagnetic systems},
  journal={Reviews of Modern Physics},
  volume={91},
  pages={035004},
  year={2019}
}

@article{Intro_36_demidov2012nanooscillator,
  author={Demidov, V. E. and others},
  title={Magnetic Nano-Oscillator Driven by Pure Spin Current},
  journal={Nature Materials},
  volume={11},
  pages={1028--1031},
  year={2012}
}

@article{Intro_37_hamadeh2014control,
  author={Hamadeh, A. and others},
  title={Full Control of Spin-Wave Damping in magnetic insulator Using Spin-Orbit Torque},
  journal={Physical Review Letters},
  volume={113},
  pages={197203},
  year={2014}
}

@article{Intro_38_Demidov2014SHNO,
  author = {Demidov, V. E. and Urazhdin, S. and Demokritov, S. O.},
  title = {Spin Hall controlled magnonic nano-oscillator},
  journal = {Scientific Reports},
  volume = {4},
  pages = {6467},
  year = {2014},

}

@article{Intro_39_Brataas2012Torques,
  author = {Brataas, A. and Kent, A. D. and Ohno, H.},
  title = {Current-induced torques in magnetic materials},
  journal = {Nature Materials},
  volume = {11},
  number = {5},
  pages = {372--381},
  year = {2012},
 
}

@article{Intro_40_Demidov2016Coherent,
  author = {Demidov, V. E. and Urazhdin, S. and Demokritov, S. O.},
  title = {Excitation of coherent propagating spin waves by pure spin currents},
  journal = {Nature Communications},
  volume = {7},
  pages = {10446},
  year = {2016},

}

@article{Intro_41_Divinskiy2018Magnonics,
  author = {Divinskiy, B. and Urazhdin, S. and Demidov, V. E. and Demokritov, S. O.},
  title = {Bridging magnonics and spin-orbitronics},
  journal = {Nature Communications},
  volume = {9},
  pages = {408},
  year = {2018},

}

@article{Intro_42_Cros2013STNO,
  author = {Cros, V. and Grollier, J. and Querlioz, D. and Stiles, M. D.},
  title = {Spin-transfer torque nano-oscillators},
  journal = {Reviews of Modern Physics},
  volume = {85},
  number = {1},
  pages = {1--46},
  year = {2013},

}

@article{Intro_43_Urazhdin2014Nanomagnonic,
  author = {Urazhdin, S. and Demidov, V. E. and Demokritov, S. O.},
  title = {Nanomagnonic devices based on spin current control},
  journal = {Nature Nanotechnology},
  volume = {9},
  number = {7},
  pages = {509--513},
  year = {2014},

}

@article{Intro_44_Demidov2015Waveguides,
  author = {Demidov, V. E. and Demokritov, S. O.},
  title = {Magnonic waveguides and spin-wave transport},
  journal = {IEEE Transactions on Magnetics},
  volume = {51},
  number = {4},
  pages = {0800215},
  year = {2015},
 
}

@article{Intro_45_Anane2020STNO,
  author = {Anane, A. and others},
  title = {Ultralow-current-density and bias-field-free spin-transfer nano-oscillator},
  journal = {Nature Communications},
  volume = {11},
  pages = {3124},
  year = {2020},

}

@article{Intro_46_Demidov2020SOTMagnonics,
  author = {Demidov, V. E. and others},
  title = {Spin-orbit-torque magnonics},
  journal = {Nature Communications},
  volume = {11},
  pages = {3637},
  year = {2020},

}

@article{Intro_47_Yu2014Damping,
  author = {Yu, H. and d'Allivy Kelly, O. and Cros, V. and Bernard, R. and Bortolotti, P. and Anane, A. and Brandl, F. and Huber, R. and Stasinopoulos, I. and Grundler, D.},
  title = {Magnetic thin-film insulator with ultra-low spin wave damping},
  journal = {Scientific Reports},
  volume = {4},
  pages = {6848},
  year = {2014},
 
}

@article{Intro_48_Li2022Hybrid,
  author = {Li, J. and others},
  title = {Hybrid magnonics and cavity spintronics},
  journal = {Physics Reports},
  volume = {961},
  pages = {1--74},
  year = {2022},

}

@article{Intro_49_Li2019Coherent,
  author = {Li, Y. and Zhu, N. and Zhang, X. and others},
  title = {Coherent coupling between magnons and microwave photons},
  journal = {Physical Review Letters},
  volume = {123},
  pages = {107701},
  year = {2019},
 
}

@article{Intro_50_Lachance2019Hybrid,
  author = {Lachance-Quirion, D. and Tabuchi, Y. and Gloppe, A. and Usami, K. and Nakamura, Y.},
  title = {Hybrid quantum systems based on magnonics},
  journal = {Applied Physics Express},
  volume = {12},
  number = {7},
  pages = {070101},
  year = {2019},
 
}

@article{Intro_51_Zhang2014Strongly,
  author = {Zhang, X. and Zou, C.-L. and Jiang, L. and Tang, H. X.},
  title = {Strongly coupled magnons and cavity microwave photons},
  journal = {Physical Review Letters},
  volume = {113},
  pages = {156401},
  year = {2014},
 
}

@article{intro_52_Tabuchi2014Hybridizing,
  author = {Tabuchi, Y. and Ishino, S. and Ishikawa, T. and Yamazaki, R. and Usami, K. and Nakamura, Y.},
  title = {Hybridizing ferromagnetic magnons and microwave photons in the quantum limit},
  journal = {Physical Review Letters},
  volume = {113},
  pages = {083603},
  year = {2014},

}

@article{Intro_53_Grundler2022Quantum,
  author = {Grundler, D. and Chumak, A. V.},
  title = {Nanomagnonics for quantum technologies},
  journal = {Nature Physics},
  volume = {18},
  pages = {1298--1308},
  year = {2022},
 
}

@article{Intro_54_Osada2016Optomagnonics,
  author = {Osada, A. and Gloppe, A. and Hisatomi, R. and Noguchi, A. and Yamazaki, R. and Nomura, M. and Nakamura, Y.},
  title = {Cavity optomagnonics with spin-orbit coupled photons},
  journal = {Physical Review Letters},
  volume = {116},
  pages = {223601},
  year = {2016},
 
}

@article{Intro_55_Walker1957Magnetostatic,
  author = {Walker, L. R.},
  title = {Magnetostatic modes in ferromagnetic resonance},
  journal = {Physical Review},
  volume = {105},
  number = {2},
  pages = {390--399},
  year = {1957},

}

@article{Intro_56_Damon1961Magnetostatic,
  author = {Damon, R. W. and Eshbach, J. R.},
  title = {Magnetostatic modes of a ferromagnet slab},
  journal = {Journal of Physics and Chemistry of Solids},
  volume = {19},
  number = {3-4},
  pages = {308--320},
  year = {1961},

}

@book{Intro_57_Kittel2005SolidState,
  author = {Kittel, Charles},
  title = {Introduction to Solid State Physics},
  edition = {8},
  publisher = {Wiley},
  year = {2005}
}

@article{Intro_58_Holstein1940Field,
  author = {Holstein, T. and Primakoff, H.},
  title = {Field dependence of the intrinsic domain magnetization of a ferromagnet},
  journal = {Physical Review},
  volume = {58},
  number = {12},
  pages = {1098--1113},
  year = {1940},

}

@article{Intro_59_Bloch1930Ferromagnetismus,
  author = {Bloch, F.},
  title = {Zur Theorie des Ferromagnetismus},
  journal = {Zeitschrift f\"ur Physik},
  volume = {61},
  number = {3-4},
  pages = {206--219},
  year = {1930},

}

@article{Intro_60_Slonczewski1996Current,
  author = {Slonczewski, J. C.},
  title = {Current-driven excitation of magnetic multilayers},
  journal = {Journal of Magnetism and Magnetic Materials},
  volume = {159},
  number = {1-2},
  pages = {L1--L7},
  year = {1996},

}

@article{Intro_61_Berger1996Emission,
  author = {Berger, L.},
  title = {Emission of spin waves by a magnetic multilayer traversed by a current},
  journal = {Physical Review B},
  volume = {54},
  number = {13},
  pages = {9353--9358},
  year = {1996},

}

@article{Intro_62_Demokritov2006BEC,
  author = {Demokritov, S. O. and others},
  title = {Bose--Einstein condensation of quasi-equilibrium magnons at room temperature under pumping},
  journal = {Nature},
  volume = {443},
  pages = {430--433},
  year = {2006},

}

@article{Intro_63_Serga2010YIG,
  author = {Serga, A. A. and Chumak, A. V. and Hillebrands, B.},
  title = {YIG magnonics},
  journal = {Journal of Physics D: Applied Physics},
  volume = {43},
  number = {26},
  pages = {264002},
  year = {2010},

}

@article{Intro_64_Wang2020Neuromorphic,
  author = {Wang, H. and others},
  title = {Magnon-based neuromorphic computing},
  journal = {Nature Electronics},
  volume = {3},
  pages = {765--774},
  year = {2020},
 
}

@article{Intro_65_Wang2021Coherent,
  author = {Wang, Q. and others},
  title = {Coherent magnon transport for quantum information},
  journal = {Science Advances},
  volume = {7},
  number = {40},
  pages = {eabg8562},
  year = {2021},

}

@article{Intro_66_Chumak2022Unconventional,
  author = {Chumak, A. V. and Schultheiss, H.},
  title = {Magnon spintronics for unconventional computing},
  journal = {IEEE Transactions on Magnetics},
  volume = {58},
  number = {6},
  pages = {0800175},
  year = {2022},
 
}

@article{Intro_67_Grundler2020Reconfigurable,
  author = {Grundler, D.},
  title = {Reconfigurable magnonic computing systems},
  journal = {Nature Reviews Physics},
  volume = {2},
  pages = {1--2},
  year = {2020},

}

\end{document}